\documentclass[aps,prb,reprint]{revtex4-2}
\DeclareUnicodeCharacter{2212}{-}
\usepackage{graphicx}
\usepackage{dcolumn}
\usepackage{bm}
\usepackage{multirow}
\usepackage{xcolor}

\usepackage{verbatim}
\usepackage{mathrsfs}
\usepackage{amsmath,amsfonts,amssymb}
\usepackage{bbm}
\usepackage{graphics}
\usepackage{amsmath}
\let\Gamma\varGamma
\let\Delta\varDelta
\let\Theta\varTheta
\let\Lambda\varLambda
\let\Xi\varXi
\let\Pi\varPi
\let\Sigma\varSigma
\let\Upsilon\varUpsilon
\let\Phi\varPhi
\let\Psi\varPsi
\let\Omega\varOmega

\usepackage{titlesec}
\titlespacing*{\section}{0pt}{*2}{*2}
\titlespacing*{\subsection}{0pt}{*2}{*2}
\titlespacing*{\subsubsection}{0pt}{*2}{*2}
\begin{document}
	\preprint{}
	\title{\textbf{Elemental Frequency-Based Supervised Classification Approach for the Search of Novel Topological Materials}}%
	\author{Zodinpuia Ralte}
	\author{Ramesh Kumar}
	\author{Mukhtiyar Singh}%
	\email{mukhtiyarsingh@dtu.ac.in; msphysik09@gmail.com}
	\affiliation{%
	Computational Quantum Materials Design (CQMD) Lab, Department of Applied Physics, Delhi Technological University, New Delhi-110042, India}%
		
	\begin{abstract}
		The machine learning based approaches efficiently solve the goal of searching the best materials candidate for the targeted properties. The search for topological materials using traditional first-principles and symmetry-based methods often require lots of computing power or is limited by the crystalline symmetries. In this study, we present frequency-based statistical descriptors for machine learning-driven topological material's classification that is independent of crystallographic symmetry of wave functions. This approach predicts the topological nature of a material based on its chemical formula. With a balanced dataset of 3910 materials, we have achieved classification accuracies of 82\% with the Support Vector Machine (SVM) model and 83\% with the Random Forest (RF) model, where both models have trained on common frequency based features. We have varified the performances of the models using $5-fold$ cross-validation approach. Further, we have validated the models on a dataset of unseen binary compounds and have efficiently identified 22 common materials using both the models. Next, we implemented the $first-principles$ approach to confirm the the topological nature of these predicted materials and found the topological signatures of Dirac, Weyl, and nodal-line semimetallic phases. Therefore, we have demonstrated that the implications of frequency-based descriptors is a practical and less complex way to find novel topological materials with certain physical post-processing filters. This approach lays the groundwork for scalable, data-driven topological property screening of complex materials. 
		\begin{description}
			\item[Keywords]
		Machine learning, topological materials, support vector machine, random forest,\\
		 $first-principles$ calculations.
		\end{description}

	\end{abstract}
	
	\maketitle
	\section{\label{sec:level1}INTRODUCTION\protect}
Materials discovery and design are essential in materials science research, which involves finding particularly promising new materials, analyzing and optimizing their attributes for various applications. Topological materials (TMs)\cite{1,2, 3, 4, 5, 6, 7, 8, 9, 10, 11, 12} have become a subject of immense interest in recent years due to their robust edge/surface states protected by different symmetries from local perturbations such as disorder, impurities, magnetic flux, etc. Since the first realization of TMs in the form of the quantum spin Hall effect \cite{13, 14}, this field has remarkably expanded to cover a number of classifications, including topological insulators (TIs) \cite{1, 2, 12, 18}, topological semimetals (TSMs),\cite{4, 5, 15, 16, 17, 21, 22, 23}, topological crystalline insulators (TCIs) \cite{3, 8, 55} etc.  Since the inception of this new class of materials, a significant and persistent investigation has revolved around the methods for ascertaining whether a specific crystalline material possesses a unique topological signature or not. The method of addressing this question has primarily relied on foundational theoretical and experimental techniques that allow trial-and-error possibilities \cite{5, 23, 24}. The first-principles computational approach and traditional experimental methods require a significant amount of time and resources \cite{25, 26, 27, 28, 29, 30, 31}. Recently developed symmetry indicators and topological quantum chemistry theories allow the analysis of a large number of TMs with low computational requirements \cite{32, 33, 34}, but despite the effectiveness of these theories, they have limitations owing to the symmetries of the crystalline materials \cite{3, 35}. For example, these theories failed to identify the topological nature of TaAs, which is a low-symmetry crystalline material and the first experimentally identified Weyl semimetal \cite{23}. Moreover, the Chern insulators and the materials having Time Reversal Symmetries (TRS), etc.\cite{14, 36, 37, 38} have robust non-trivial topological phases even when crystalline symmetries vanish. Therefore, we require novel approaches that overcome the aforementioned hurdles, highly accurate and easily executable for the exploration of TMs in condensed matter physics. 
The machine learning (ML) approach has become one of the adopted methods for screening materials with desired properties \cite{39, 40}. This approach is computationally cheaper, data-driven, with high accuracy and precision, making it an indispensable tool in material discovery \cite{34, 41, 42, 43, 44}. Various reports have been published in recent times that have predicted electronic and structural properties using regression and classification ML models based on chemical composition and crystal structures  \cite{35, 45, 46, 47, 48}. Recent advancements include graph-based methods such as crystal graph convolution neural networks (CGCNNs), which have achieved significant success in the prediction and identification of topological phases \cite{49}. A machine-learned chemical descriptor has also been introduced by Ma $et$ $al.$, thus contributing to the advancements and growing implementation of ML in TMs \cite{50}. The frequency of an atom or an element in a given set of data is an important indicator of what type of elements are more feasible to form TMs. In this study, we use a frequency-based formulation to create a descriptor that can predict TMs with acceptable precision and accuracy. This formulation is straightforward, amenable to calculation, and given materials can be categorized with an accuracy of more than 80\% in topological trivial or non-trivial classes. This formulation is independent of the crystal symmetries of the systems and can be implemented on all types of materials.  We have integrated our study using ML algorithms, $i.e.$, Support Vector Machine (SVM) and Random Forest (RF), followed by $first-principles$ calculations, for the sake of the verification of the predicted materials. This work aims to improve the accuracy and efficiency of ML models to classify materials for their non-trivial topological phases.
	\section{\label{sec:level2}METHODOLOGY\protect}
	We adopted a statistical frequency-based method to construct features \cite{51}, unlike the conventional approaches that rely on intrinsic material properties for the classification of TMs. The frequency-based method and ML models can capture a pattern in the formation of TMs based on the occurrence of elements across different topological classes. While the frequent occurrence of elements that are common in certain topological phases doesn't always guarantee that such phases will form when those elements are present, it is well known statistically that materials are more likely to create a topological phase if those elements are common in more than one topological phase. With this insight, we direct our study and methods for feature construction.
	\subsection{\label{sec:level3}Data preprocessing\protect}
	The data for our study were collected from a previous literature repository provided by Andrew $et al.$ \cite{50}. In addition, we have included 33 literature-based TI materials to maintain the uniformity of the elements and broaden the perspective of our study \cite{34}. We segregated the data into three categories, where the first group ($\mathrm{G_1}$) comprises 6098 trivial materials and the second group ($\mathrm{G_2}$) contains 1376 non-trivial TMs. Together, the first and second groups of materials make up a dataset from which the features for our models were calculated. The third group ($\mathrm{G_3}$) consists of 3371 raw materials from which the binary materials were further filtered out for the classification. We have calculated the frequency (number of occurrences) of elements for the trivial group $\mathrm{G_1}$ and the non-trivial group $\mathrm{G_2}$, denoted by $\mathit{F_x^t}$ and $\mathit{F_x^{nt}}$ respectively, for each $x$ element. Our dataset consisted of the materials, which were made up of 74 elements, including radioactive elements and rare earth elements. Due to a larger number of trivial materials compared to non-trivial materials in our datasets, there is a bias in the frequency of elements in the trivial dataset, and as a result, for each $x$ element, $\mathit{F_x^t\gg F_x^{nt}}$. To overcome this bias, we introduced a rescaling approach (Eq. (1)) that normalizes the $\mathit{F_x^t}$ and rescaled it relative to $\mathit{F_x^{nt}}$, where the rescaled frequencies are given by $\mathit{F_x^{rt}}$,
		\begin{equation}
                      	F_x^{rt}=\frac{F_x^t}{\sum_xF_x^t}\times\sum_xF_x^{nt}.
        \end{equation}
	This equation kept the relative proportions of trivial frequencies but scaled them to match the total volume of non-trivial frequencies. The various terms included in the equation are defined in Table I. 
\begin{table*}[t]  
	\centering
	\caption{Definitions of statistical terms used in frequency-based feature construction.}
	\label{tab:freq_terms}
	\begin{tabular}{|p{4cm}|p{11cm}|}
		\hline
		\textbf{Terms} & \textbf{Descriptions} \\
		\hline
		Input Term ($F_x^t$) & Frequency of element $x$ in group $\mathrm{G_1}$. \\
		\hline
		Normalization by $\sum_x F_x^t$ & The term $\frac{F_x^t}{\sum_x F_x^t}$ calculates the proportion of $F_x^t$ relative to the total sum of frequencies in the original set. This normalizes $F_x^t$ to a fractional value, ensuring the sum of all fractions equals 1. \\
		\hline
		Scaling Back with $\sum_x F_x^{nt}$ & The normalized fraction is multiplied by the sum of another set of values $\sum_x F_x^{nt}$, effectively rescaling it to a new total while preserving the original proportionality. \\
		\hline
	\end{tabular}
\end{table*}
	We have also calculated the difference between the frequency of elements as	    
		\begin{equation}
		\Delta F=F_x^{rt}-F_x^{nt}
	\end{equation}                                        
	These three quantities in Eq. (2) for each respective element are used to generate the three features for our ML models [refer to section 1.1 and Table S1 in supplementary information] \cite{20}. We generate these features by a weighted sum formula, giving three feature quantities $\mathit{f_1}$, $\mathit{f_2}$, and $\mathit{f_3}$ calculated from the quantities $\mathit{F_x^{nt}}$, $\mathit{F_x^{rt}}$, and $\mathit{\Delta F}$ respectively.
		\begin{equation}
		f_1=\sum_x(p_x\times F_x^{nt})
	\end{equation}
	\begin{equation}
		f_2=\sum_x(p_x\times F_x^{rt})
	\end{equation}
		\begin{equation}
		f_3=\sum_x(p_x\times \Delta F)
	\end{equation}
	Where $\mathit{p_x}$ is the proportion of the $x$ element present in the materials. We have prepared a dataset to train the ML model by calculating the three features for each material using Eqs. (3) to (5). The number of materials in $\mathrm{G_1}$ is larger compared to that in $\mathrm{G_2}$, which will create a bias towards the trivial class of materials due to class imbalance. To address this class imbalance and to reduce this bias between trivial and non-trivial materials, we employed the random undersampling of the majority class ($\mathrm{G_1}$) and some materials in the minority class ($\mathrm{G_2}$) as a consequence of the undersampling, which prevents skewed classification boundaries. Hence, this resulted in the inclusion of 2655 trivial (label 0) materials from $\mathrm{G_1}$ and 1255 non-trivial (label 1) materials in the balanced training and testing datasets ($\mathrm{T_1}$). The remaining non-trivial materials were not included in this dataset due to the undersampling strategy. The impact of this reduction was assessed, and the models retained strong performance across evaluation metrics. We have also calculated the features for $\mathrm{G_3}$ materials using Eqs. (3) to (5) and extracted binary materials only to create a dataset ($\mathrm{T_2}$) to implement the trained ML models.
	\subsection{\label{sec:level3}Machine Learning Implementation\protect}
	To test our features for a more complex applicability, we employed both Support Vector Machine (SVM) and Random Forest (RF) classifiers, where we used a linear kernel for the SVM, making it a linear classification model, whereas the RF is a tree-based classifier and non-linear in nature. We have split the $\mathrm{T_1}$ into an 80:20 proportion for training and testing. The trained models were implemented on the dataset $\mathrm{T_2}$, which contains 1613 binary materials. The ML models were trained on a dataset containing materials of different formula units to capture the complex compositional dependencies and to broaden the model’s chemical space coverage; however, these were implemented on the unseen dataset of only binary materials. Binary materials are simpler in composition, allowing us to assess the model’s ability to generalize from complex systems, which serves as a robust validation strategy for our frequency-based approach. By limiting it to binary materials, we significantly reduced the prediction space, thereby minimizing false predictions. However, our method is not limited to binary materials only and can be applied to any type of materials. In the final stage of prediction, we included the absolute electronegativity difference ($\mathit{\Delta E}$) between the two elements as a filter in the binary materials to further increase the likelihood of the prediction. Although $\mathit{\Delta E}$ was not a part of the ML models' training space, it is used as a post-prediction physical filter parameter. A smaller value of $\mathit{\Delta E}$ tends to favour covalent bonding, which is more susceptible to band inversion. This interplay has been captured in a predictive criterion proposed by Cao et al. \cite{52}.  This notion also extends to semimetals, where a smaller $\mathit{\Delta E}$ facilitates orbital overlap and hybridization, which in turn can enable symmetry-protected band crossings necessary for semimetallic topological phases. The SVM and RF models have been optimized to predict the topological phase of materials with more than 90\% normalized prediction probability. Moreover, the post-processing filter, $\mathit{\Delta E}$ was set at less than 1 and 0.4, respectively, for these models. The choice of different thresholds of $\mathit{\Delta E}$ for the SVM and RF models is driven by their distinct prediction behaviours, particularly their confidence in predictions and tendency to overfit. This tighter filter ensures that the materials with a stronger covalent character are predicted as non-trivial TMs.	
		\begin{figure}
		\centering
		\includegraphics[width=1\linewidth]{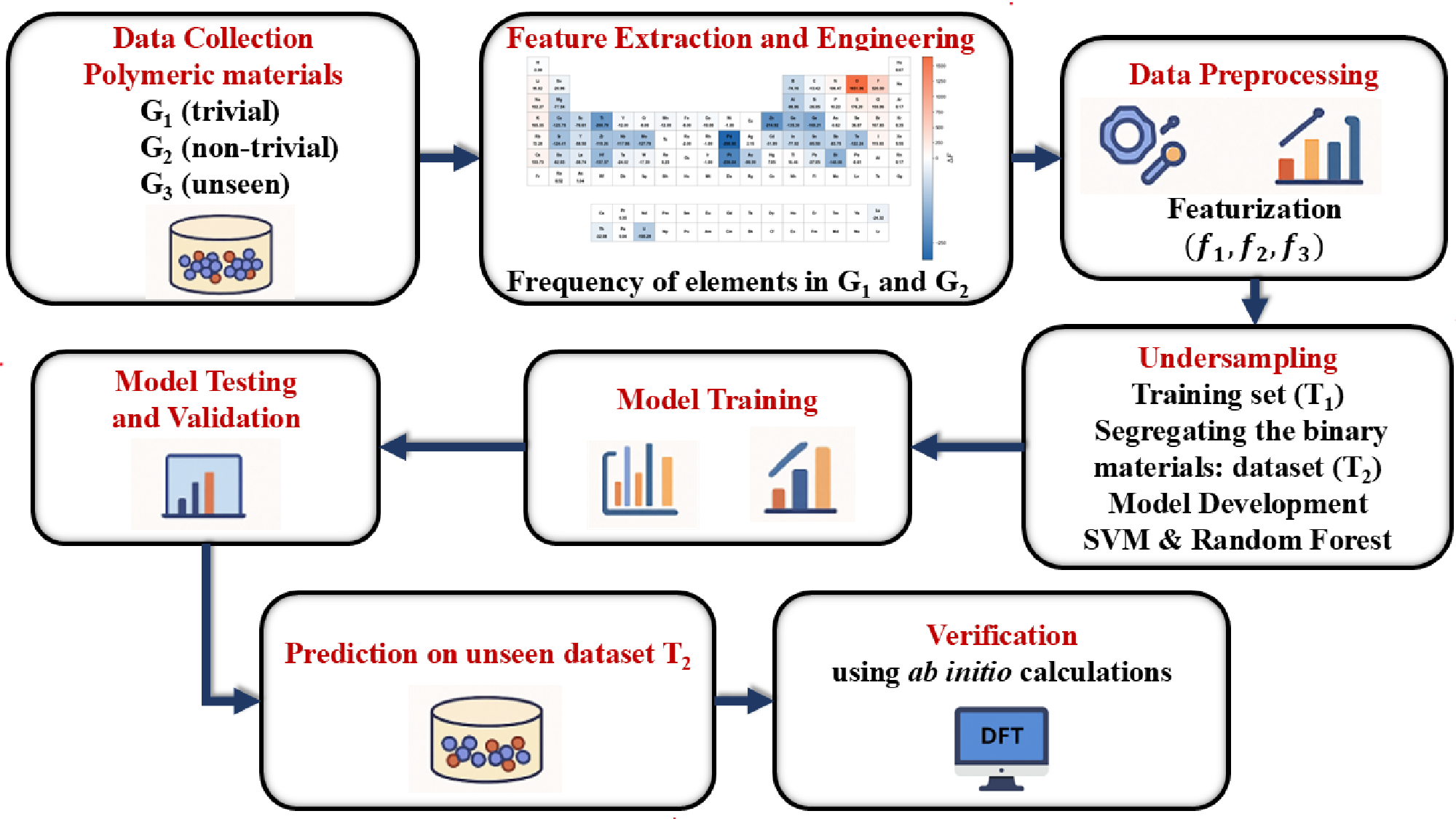}
		\caption{The workflow of statistical frequency-based feature engineering approach for training and testing of ML models, and verification of predicted materials using $first-principles$ calculations.}
		\label{fig:enter-label}
	\end{figure}
			\subsection{\label{sec:level3}\textit{First-principles} calculations\protect}
		The electronic structure calculations based on the density functional theory (DFT) \cite{53} were performed using the projector augmented-wave (PAW) \cite{54} method as implemented in the Vienna ab initio Simulation Package (VASP) \cite{56} code. The standard PAW potentials, preferably suggested in the VASP code, were used for all the elements for these calculations. The exchange and correlation energy were obtained with the generalized gradient approximation (GGA) of Perdew Burke Ernzerhof (PBE) functional \cite{60}, with the inclusion of SOC in the self-consistent field calculations \cite{61}. The total energies convergence criterion of $\mathrm{10^{-6}}$ eV was adopted along with a fine gamma-centered k mesh. The plane-wave cut-off energies were set at 1.5 times the maximum energies available in the PAW potentials. 
	\section{\label{sec:level3}RESULTS AND DISCUSSION\protect}
	\subsection{\label{sec:level3}Training and Testing\protect}
	The input dataset with non-trivial and trivial datapoints is the primary requirement for creating the classification models (SVM and RF). Our dataset ($\mathrm{T_1}$) contains the trivial materials from group $\mathrm{G_1}$ and non-trivial materials from $\mathrm{G_2}$ after the undersampling process. The satisfactory performance of both models can be observed from the confusion matrix as shown in Fig. 2 (a, b). The confusion matrix characterizes the materials with the four parameters, namely True Positive, True Negative, False Positive, and False Negative. It is observed that both models have similar results in correctly predicting the two classes of materials, but the RF model, which is an ensemble model, has a slight edge in its prediction capability, correctly identifying more topological classes of materials. This trend is also established with a slightly higher value of the Area Under the Curve (AUC) of the RF model calculated using the Receiver Operating Characteristic (ROC) curve plotted between the true positive and false positive rates (Fig. 2 (c, d)). These two metrics offer an insight into the two models' ability to differentiate between the two classes of materials, and their values showed a remarkable reliability for the features we have created.
		\begin{figure}
		\centering
		\includegraphics[width=1\linewidth]{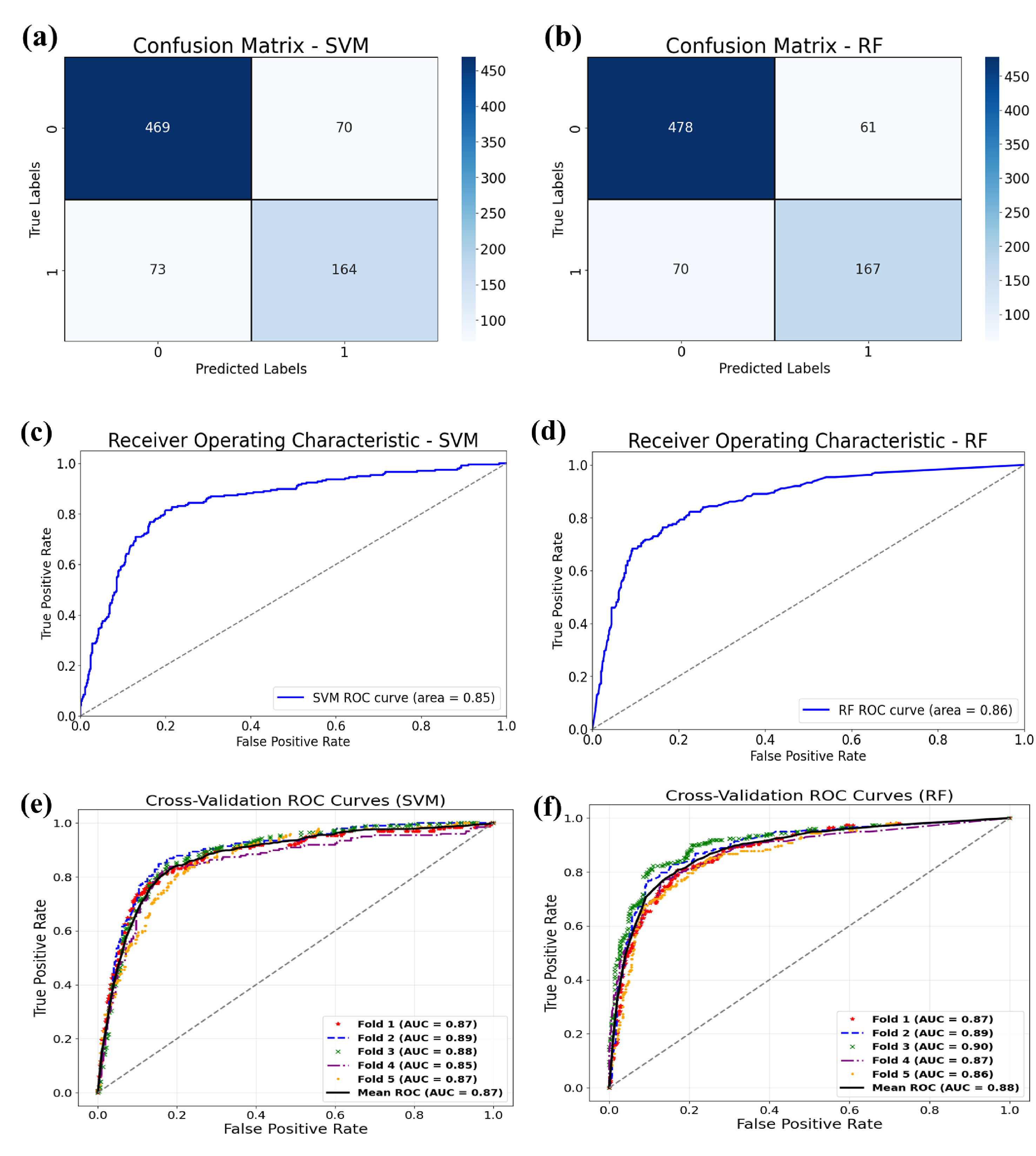}
		\caption{(a, b) Metrics of classification showing Confusion matrices and (c, d) Receiver Operating Characteristic (ROC) - Area Under the Curve (AUC) for the models SVM and RF, respectively. (e, f) ROC-AUC curves for the $5-fold$ cross-validation of SVM and RF models.}
		\label{fig:enter-label}
	\end{figure}
	The number of trivial materials being more in the testing set can be seen from the first row of the two matrices, indicating the confidence of the models to distinguish trivial materials and at the same time have a high confidence to classify non-trivial materials, as can be seen in the second row.
	To further verify the reliability and interpretability of these models, we have calculated other metrics of classifications and analyzed their performance using Precision, Recall, F1 score, and the Area Under the Curve (AUC) parameters as given in Table II, for the target materials. These parameters can be calculated using the following equations,
		\begin{equation}
	Precision=\frac{True\ positive}{True\ positive+False\ positive}
	\end{equation}
		\begin{equation}
		Recall=\frac{True\ positive}{True\ positive+False\ negative}
	\end{equation}
		\begin{equation}
		F1=2\frac{Precision\times Recall}{Precision+Recall}
	\end{equation}
	\begin{table}[h]
		\centering
		\caption{Metrics of classification for the models SVM and RF, showing Precision, Recall, F1, AUC, and Accuracy of the classifiers on the test case from $\mathrm{T_1}$.}
		\begin{tabular}{|c|c|c|c|c|c|} \hline 
			ML Model& Precision& Recall& F1 & AUC & Accuracy (\%)\\ \hline 
			SVM& 0.70& 0.69& 0.70 & 0.85 & 82\\ \hline 
		    RF& 0.73& 0.70& 0.72 & 0.86 & 83 \\ \hline 
		\end{tabular}
		\label{tab:my_label}
	\end{table}
	The numerical results of the RF model are relatively better in comparison to the SVM model due to its nature as an ensemble and its ability to capture complex patterns.  The AUC scores for the SVM and RF models are 0.85 and 0.86, respectively, which shows that they are very good at telling different classes apart, indicating that the classifiers are reliable and useful for further analysis, especially when combined with methods that look at feature importance or use probability for filtering. The accuracy of both models is 82\% and 83\%, respectively, which is almost the same and indicates that the features we have engineered are robust and work well for both types of classifiers. 
		\subsubsection{\label{sec:level3}Performance analysis of the models\protect}
		For the SVM model, the learned weight vector w = [0.3667, -1.8711, -2.2649] corresponds to the features  $\mathit{f_1}$,  $\mathit{f_2}$, and  $\mathit{f_3}$, respectively. The positive coefficient for  $\mathit{f_1}$ suggests that there is a direct, though weak, link between  $\mathit{f_1}$ and the non-trivial class. The strong negative coefficients for  $\mathit{f_2}$ and  $\mathit{f_3}$ show that higher values (toward positive values) of these features are linked to the negative (trivial) class. On the other hand, a lower value for these features suggests that the class material is more likely to be non-trivial. This means that  $\mathit{f_2}$ and $\mathit{f_3}$ have a lot of information that helps differentiate between classes, and are the most important factors in the classification decision. Feature $\mathit{f_1}$ doesn't contribute as much, which could mean that it doesn't have as much predictive power or that it is partially redundant with other features. The $\mathit{f_3}$ value has the greatest ability to tell the difference between the materials. This is also in accordance with our findings that all the materials classified by the SVM model had a negative $\mathit{f_3}$ value. 
		The model uses 1657 support vectors, which are almost evenly distributed between the two classes (829 for trivial and 828 for non-trivial). A high number of support vectors usually means that the decision boundary is complicated, which is needed to pick up on small differences in the data distribution. The fact that there are almost equal numbers of support vectors from both classes shows that the SVM balances class representation when it sets the margin. The geometric margin is about 0.338, which is the distance between the decision boundary and the support vectors that are closest to it. In general, larger margins are better because they help with generalization. However, in this case, the margin is average, which means that there is a trade-off between margin size and classification accuracy. There are 405 slack variables, which are instances that break the margin rules. This shows that the classes are not completely linearly separable, which means that the feature space has some noise or overlap. This explains why the model doesn't get perfect recall on the non-trivial class of materials. On the other hand, the RF model is very useful for complex datasets because it can naturally find non-linear relationships in the data. In addition, it has a built-in way to rank the importance of different features. As the features ($\mathit{f_1}$,  $\mathit{f_2}$ , and $\mathit{f_3}$) used in this study come from the differences between the frequencies of the elements, it is hard to understand their interactions.  Here, the strength of the RF model comes from the fact that it can implicitly consider these interactions across its many decision trees. Hence, the feature importance scores for $\mathit{f_1}$,  $\mathit{f_2}$ , and $\mathit{f_3}$, come out as 0.2192, 0.4109, and 0.3699, respectively, indicating that the feature  $\mathit{f_2}$  is the most important in differentiating the materials. 
			\subsubsection{\label{sec:level3}Cross Validation\protect}
		Cross-validation is a key part of ML analysis, as it permits the removal of reliance on how the sample is divided into testing and training datasets for performance estimation. The cross-validation helps to discover and fix the problem of overfitting and underfitting, and it is a better way to estimate the generalization error of the training-testing split, as it tests the model on more than one validation dataset. This method is used to fine-tune the hyperparameters so they can pick the ones that work best for generalization. Cross-validation also helps make fair comparisons between different models by testing each one on several subsets of data under the same conditions. The complete dataset, consisting of $n$ elements, is randomly divided into $p$ subsets, each containing $n/p$ elements, referred to as folds, as shown in Table III. Here, we have carefully tested the performance of the SVM and RF models by using  $\mathit{five-fold}$ cross-validation (FIG. 2 (e, f)), and the AUC scores of each fold, as well as the mean AUC score, are shown in Table III. 
	
	\begin{table*}[t]  
		\centering
		\caption{The AUC scores of SVM and RF models, as well as the average AUC value in the $five-fold$ cross-validation.}
		\label{tab:auc_scores}
		\begin{tabular}{|c|c|c|c|c|c|c|}
			\hline
			\textbf{ML Model} & 	\textbf{Fold 1} & 	\textbf{Fold 2} & 	\textbf{Fold 3} & 	\textbf{Fold 4} & 	\textbf{Fold 5} & 	\textbf{Mean Cross-validation} \\ \hline
			SVM (AUC) & 0.87 & 0.89 & 0.88 & 0.85 & 0.87 & 0.87 \\ \hline
			RF (AUC)  & 0.87 & 0.89 & 0.91 & 0.87 & 0.86 & 0.88 \\ \hline
		\end{tabular}
	\end{table*}
	
	The AUC scores for the SVM and RF models go from 0.85 to 0.89 and 0.86 to 0.91, with average scores of 0.87 and 0.88, respectively, indicating that the models perform well and reliably across different data sets in the $five-fold$ cross-validation analysis. Both models have a small range of variability, which confirms the strong and insensitive nature of models to certain data partitions. The AUC scores for the SVM and RF models on the test sets are 0.85 and 0.86, which are lower than the average score from cross-validation, showing how well the models performed on a separate test set, different from the cross-validation folds. The AUC score in the test performance and $five-fold$ cross-validation is slightly higher for RF than the SVM model and is slightly superior in distinguishing between the non-trivial and trivial classes across various classification thresholds. 
		\subsection{\label{sec:level3}Analysis of data in $\mathrm{T_1}$\protect}
		To support our frequency-based feature approach, we have analyzed the elemental data in the dataset $\mathrm{T_1}$ and calculated the elemental features $\mathit{f_1}$, $\mathit{f_2}$ using Eqs. (3) and (4), whereas the frequency difference $\mathit{\Delta F}$ and feature $\mathit{f_3}$ have been calculated using Eqs. (2) and (5), respectively. We have observed that the $\mathit{\Delta F}$ values are positive for the alkali metals group (with the exception of hydrogen), whereas the halogens have highly positive $\mathit{\Delta F}$ values in the $p-block$ elements. Moreover, the highly electronegative elements, such as Oxygen, have an unusually high value of $\mathit{\Delta F}$, which reflects their widespread occurrence across a varied range of materials in the dataset. The elementary distribution of the rest of the data for $\mathit{\Delta F}$ across the periodic table can be seen in the heatmap shown in Fig. 3. Among the elements with high negative $\mathit{\Delta F}$ values, Pd, Pt, Ti, Hf, Zr, U, Au, and Sc occur in more than 50\% of the materials containing these elements and are labelled as topologically non-trivial. This observation and analysis imply a higher probability that materials containing these elements possess topological phases. Furthermore, the other elements with negative $\mathit{\Delta F}$ values generally have a 30\% to 50\% chance of being in a material of a topological phase. Hence, it implies, statistically, that there is indeed a chance to form topological phases of materials when the elements with $\mathit{\Delta F <0}$ are present in a material. Moreover, the elements with $\mathit{\Delta F>0}$ have shown a less than 30\% chance of contributing towards this phase. These conclusions are in line with the observation made in the dataset, where most elements with ionic characters are less likely to form topological phases due to their high electronegativity values. This reinforces the significance of the $\mathit{\Delta F}$ metric as a discriminative feature in identifying potential topological materials and supports its integration into our ML framework. Table S2 in the supplementary information \cite{20} shows the percentage occurrence of the elements in TMs.
			\begin{figure}
			\centering
			\includegraphics[width=1\linewidth]{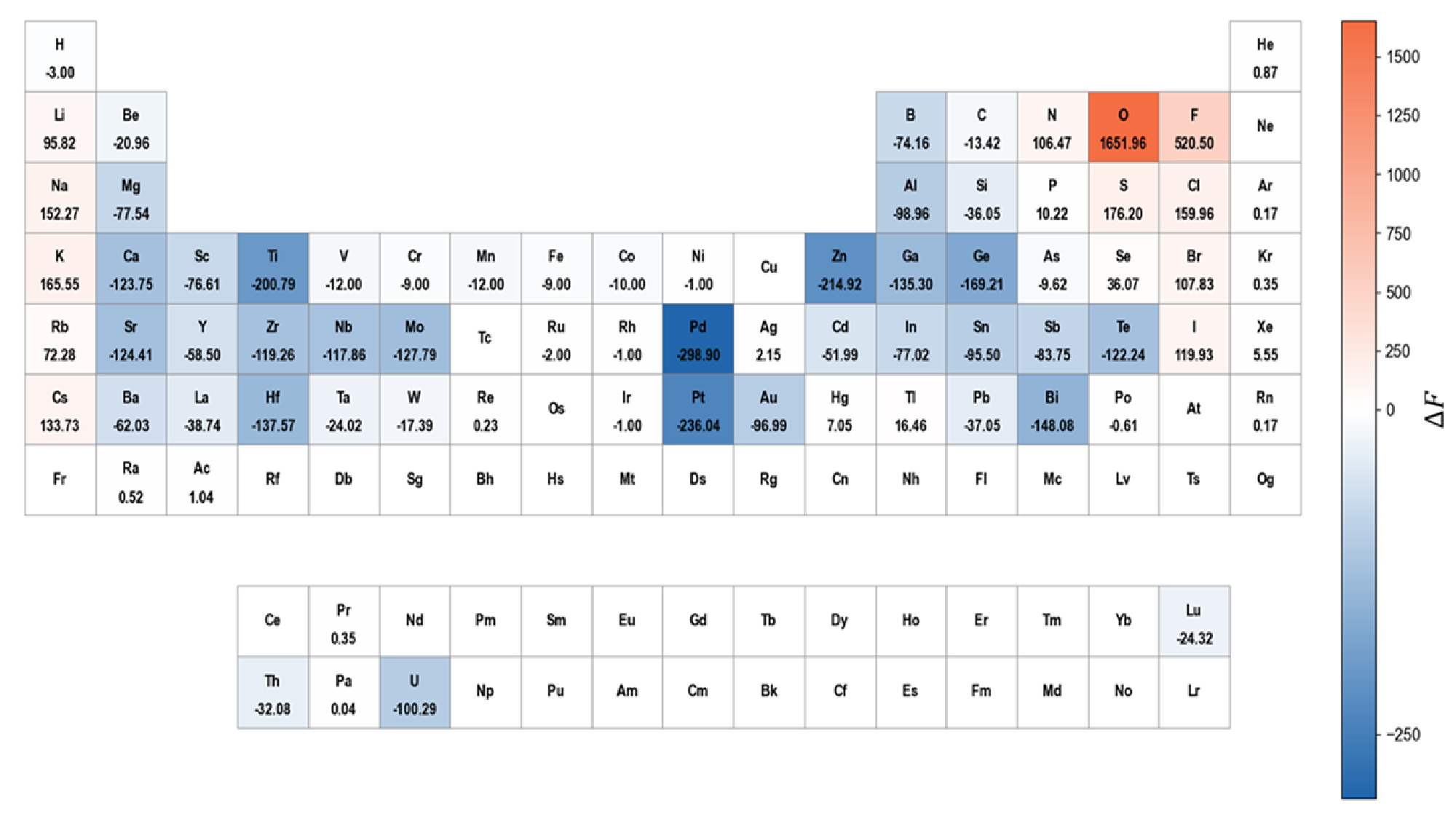}
			\caption{Heatmap of the difference in the frequency of elements from $\mathrm{G_1}$ and $\mathrm{G_2}$ given in Eq. (2).}
			\label{fig:enter-label}
		\end{figure}
		The single-element $\mathit{\Delta F}$ analysis, in the preceding discussion, has extended our investigation to consider two-element combinations to further validate the predictive capability of our frequency-based approach and to verify the aforementioned analysis of the dataset $\mathrm{T_1}$. Table S3 in the supplementary information \cite{20} lists the various pairs of two-element combinations, the total number of materials containing that pair, the number of TMs having that pair, along with their percentage of occurrence. Our empirical analysis of the dataset  $\mathrm{T_1}$ in Table S3 reveals that a certain combination of element pairs shows a strong correlation with topological phases such as Pd–Sr and C–Nb, Se–U, Ga–Pt, Pd–Si, and Te–U, etc., suggesting that the co-occurrence of specific elements enhances the likelihood of band inversion and non-trivial topology in the materials. The repeated presence of some elements and pairwise elemental interactions offers additional predictive value, especially for binary materials, and reinforces the reliability of $\mathit{\Delta F}$ as a descriptor when screening for candidate materials. Furthermore, several of the high-frequency combinations exist with transition metals, heavy $p-block$ elements, alkaline earth metals, etc., which typically have high atomic numbers, spin-orbit coupling, and favorable electronegativity profiles. This two-element analysis also supports the filtering strategy implemented in post-prediction of ML models, where elemental electronegativity and their co-occurrence trends are used to refine model outputs. We have increased confidence in our statistical approach by cross-referencing frequently co-occurring element pairs with  $\mathit{\Delta F<0}$ in the selected candidates upon implementing our trained model on  $\mathrm{T_2}$. 
		\subsection{\label{sec:level3}Classification of unseen data ($\mathrm{T_2}$)\protect}
		In the preceding section, we have trained and tested it on the labelled data available in the dataset $\mathrm{T_1}$; now, we have implemented both models on the unseen data, which is assigned in the dataset $\mathrm{T_2}$. It should be noted that the dataset $\mathrm{T_2}$ contains the binary combinations of the elements only. As per the observations, the existence of a topological phase depends on the negative value of the $\mathit{\Delta F}$; hence, using Eq. (5), $\mathit{f_3}$ is also likely to be negative, which concludes that all the predicted topological materials possess $\mathit{f_3<0}$. The non-trivial topological phase using SVM and RF models has been predicted, respectively, in 84 and 86 materials out of a 1613 binary dataset ($\mathrm{T_2}$). We have performed the $first-principles$ calculations of 30 (Table IV and Table V) selected materials out of 84 and 86 topologically non-trivial materials predicted by the SVM and RF models, respectively. The $first-principles$ calculations have shown that these materials have topological semimetallic phases (i.e., DSMs, WSMs, and NLSMs). The crystal structures and band structures of the some selected materials predicted by SVM and RF models are shown in Fig. 4, 5 and Fig. 6, 7, respectively. The corresponding k path selected for band structure calculations in the first Brillouin zone are shown in Fig. S3 and Fig. S4. For the band structure analysis of rest of the systems, Fig. S6 and Fig. S7 can be referred in supplementary information\cite{20}. 
\begin{table}[htbp]
			\centering
			\caption{The identified topological phases of the predicted materials with SVM model using $first-principles$ calculations.}
			\begin{tabular}{|c|c|c|c|c|c|}
				\hline
				\textbf{Material} & \textbf{SG No.} & \textbf{Phase} & \textbf{Material} & \textbf{SG No.} & \textbf{Phase} \\
				\hline
				Al$_3$Pd$_5$ & 55  & DSM   & ZrPd     & 221 & WSM \\
				TaPd$_2$     & 71  & DSM   & TiPt     & 221 & WSM \\
				HfPt$_3$     & 221 & DSM   & Ga$_5$Pd$_{13}$ & 12  & WSM \\
				ZrPd$_3$     & 221 & DNLSM & Ti$_3$Pd & 223 & WSM \\
				ZrPt$_3$     & 221 & DSM   & Ti$_3$Pt & 223 & WSM \\
				TlPd$_3$     & 139 & DNLSM & TiPt$_3$ & 221 & WSM \\
				AlPd$_2$     & 62  & DNLSM & TiPd     & 221 & WSM \\
				YPd$_3$      & 221 & WSM   & TaPd$_3$ & 139 & WSM \\
				NbPd$_3$     & 139 & WSM   & Ga$_3$Pd$_5$ & 55  & WSM \\
				TiPd         & 123 & WSM   & GaPt$_3$ & 221 & WSM \\
				NbPd$_2$     & 71  & WSM   & HfPd     & 221 & WSM \\
				TiPd$_3$     & 221 & WSM   & MgPd$_3$ & 221 & WSM \\
				LuPd$_3$     & 221 & WSM   & NbPt$_3$ & 59  & WSM \\
				InPd$_3$     & 221 & WSM   & Ga$_3$Pd$_7$ & 12  & WSM \\
				BiPd         & 194 & WSM   & InPd$_2$ & 62  & WSM \\
				\hline
			\end{tabular}
			\label{tab:svm_tm_materials}
		\end{table}
\begin{figure}
			\centering
			\includegraphics[width=0.6\linewidth]{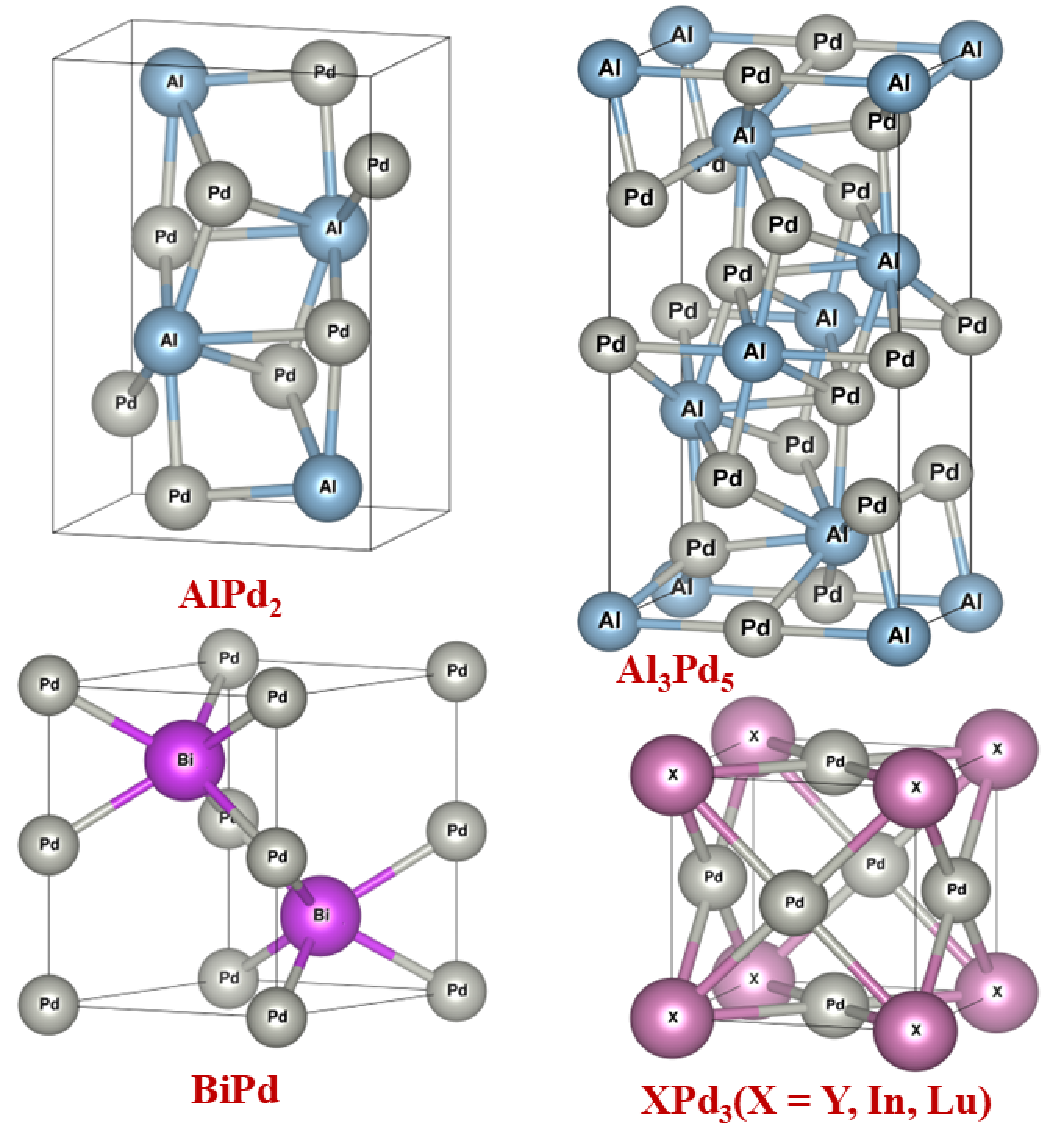}
			\caption{The crystal structures of some selected materials from the predicted SVM datapoints.}
			\label{fig:enter-label}
		\end{figure}
\begin{figure}
			\centering
			\includegraphics[width=1\linewidth]{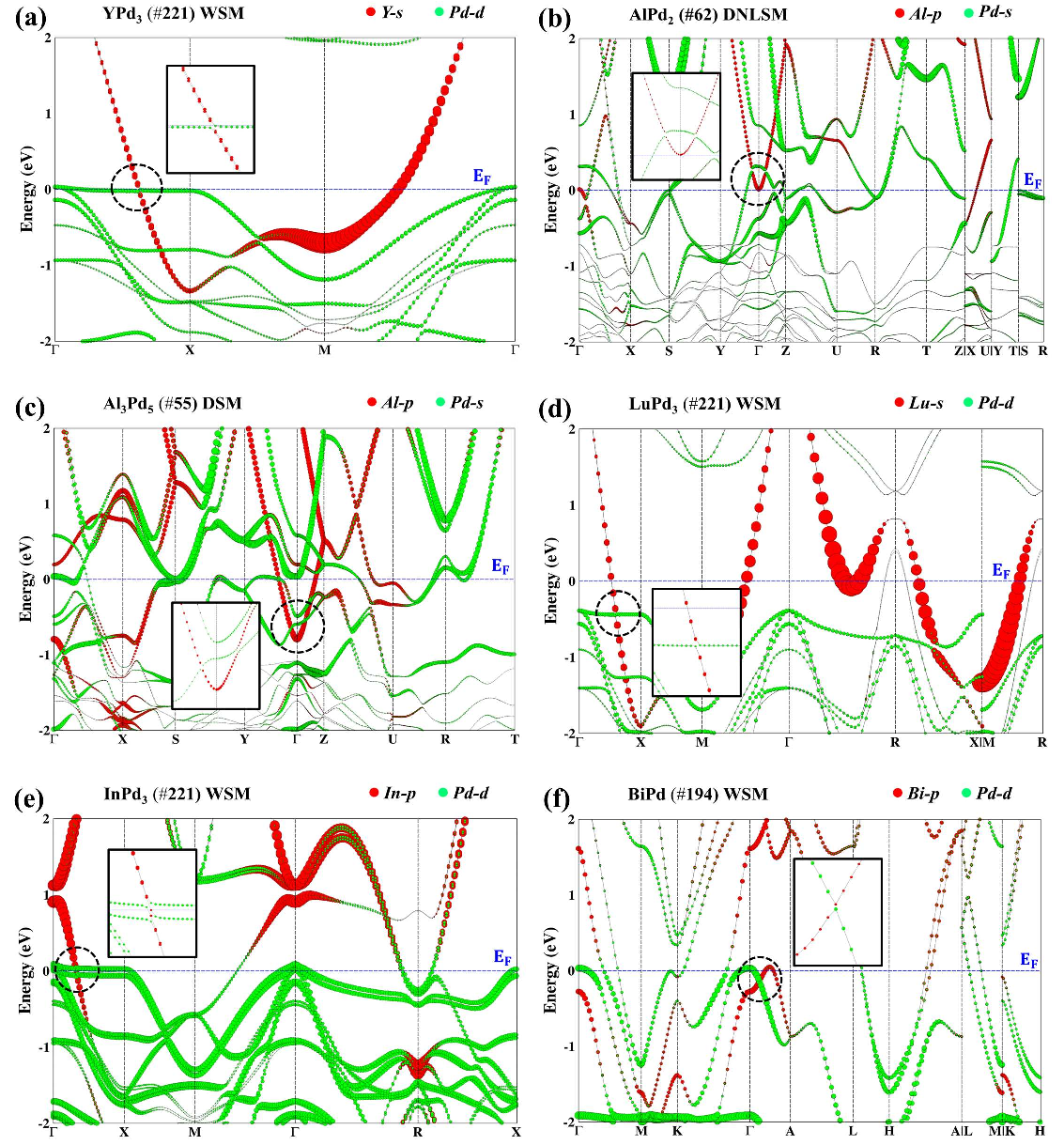}
			\caption{The projected band structures of some selected materials , $i.e.$, (a) YPd$_3$, (b) AlPd$_2$, (c) Al$_3$Pd$_5$, (d) LuPd$_3$, (e) InPd$_3$, (f) BiPd, from the predicted SVM datapoints.}
			\label{fig:enter-label}
		\end{figure}
\begin{table}[htbp]
			\centering
			\caption{The identified topological phases of the predicted materials with RF model using $first-principles$ calculations.}
			\begin{tabular}{|c|c|c|c|c|c|}
				\hline
				\textbf{Material} & \textbf{SG No.} & \textbf{Phase} & \textbf{Material} & \textbf{SG No.} & \textbf{Phase} \\
				\hline
				TePt       & 194 & DSM   & ScBe$_5$   & 191 & WSM \\
				ScMg       & 221 & DSM   & PdAu$_3$   & 221 & WSM \\
				TiAl       & 123 & DSM   & SbPt       & 194 & WSM \\
				Be$_2$Nb$_3$ & 127 & DNLSM & ScCd       & 221 & WSM \\
				GaMo$_3$   & 223 & DNLSM & Nb$_3$Al   & 223 & WSM \\
				NbGa$_3$   & 139 & DNLSM & Nb$_3$Ga   & 223 & WSM \\
				Th$_2$Zn   & 140 & DSM   & SiMo$_3$   & 223 & WSM \\
				BiB        & 216 & DSM   & Nb$_3$Tl   & 223 & WSM \\
				HfAl$_3$   & 221 & DSM   & Be$_2$Nb   & 227 & WSM \\
				In$_2$Bi   & 194 & DNLSM & Cd$_3$Zr   & 123 & WSM \\
				TiAg       & 123 & DNLSM & Y$_2$Al    & 62  & DSM \\
				Ti$_2$Re   & 64  & DNLSM & HfBe$_5$   & 191 & WSM \\
				HfBe$_2$   & 191 & WSM   & TiAg       & 129 & WSM \\
				ZrBe$_2$   & 191 & WSM   & Th$_3$Al$_2$ & 127 & WSM \\
				YMg        & 221 & WSM   & MoC        & 194 & WSM \\
				\hline
			\end{tabular}
			\label{tab:rf_tm_materials}
		\end{table}
\begin{figure}
			\centering
			\includegraphics[width=0.6\linewidth]{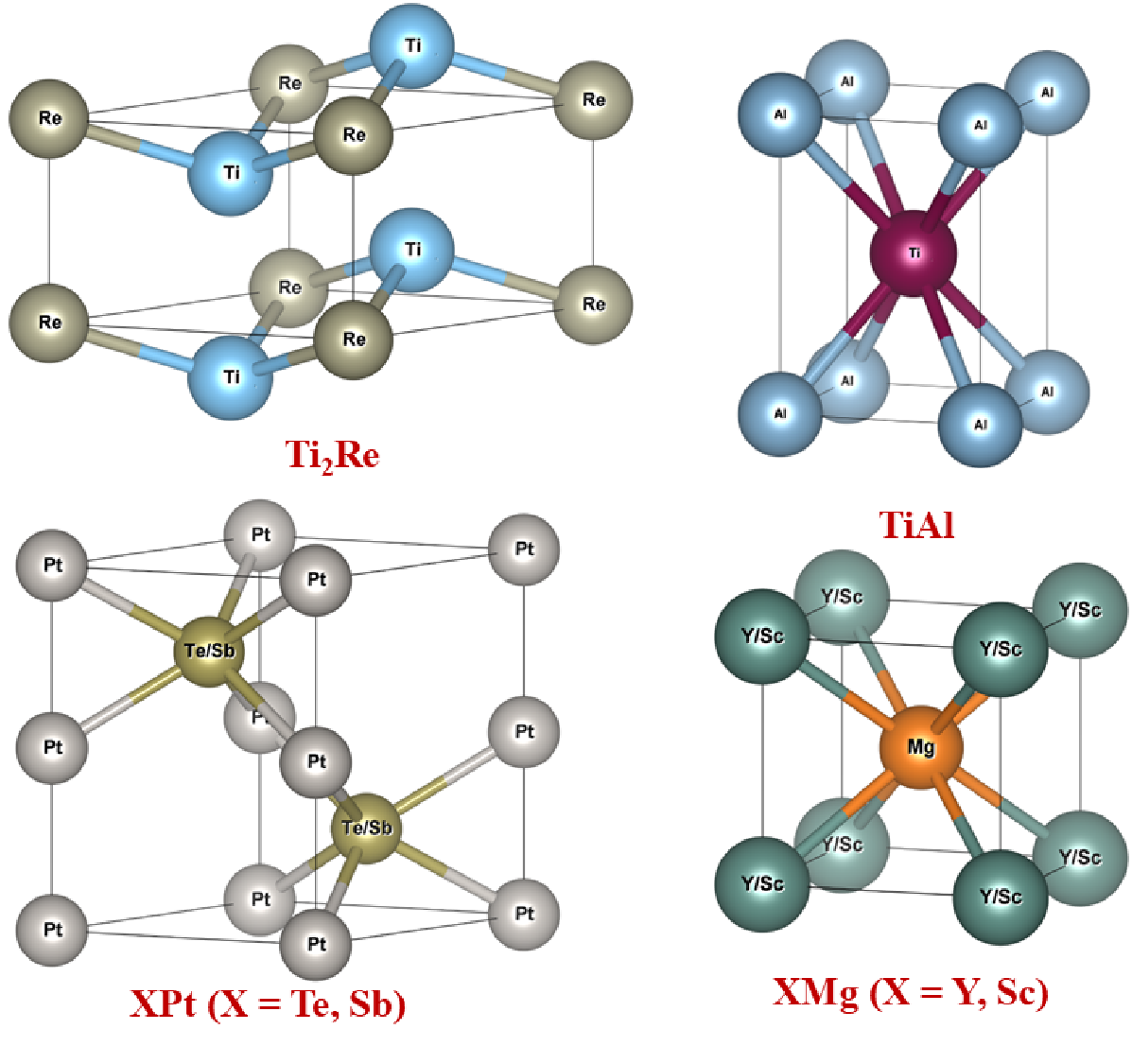}
			\caption{The crystal structures of some selected materials from the predicted RF datapoints.}
			\label{fig:enter-label}
		\end{figure}
		
\begin{figure}
	\centering
	\includegraphics[width=1\linewidth]{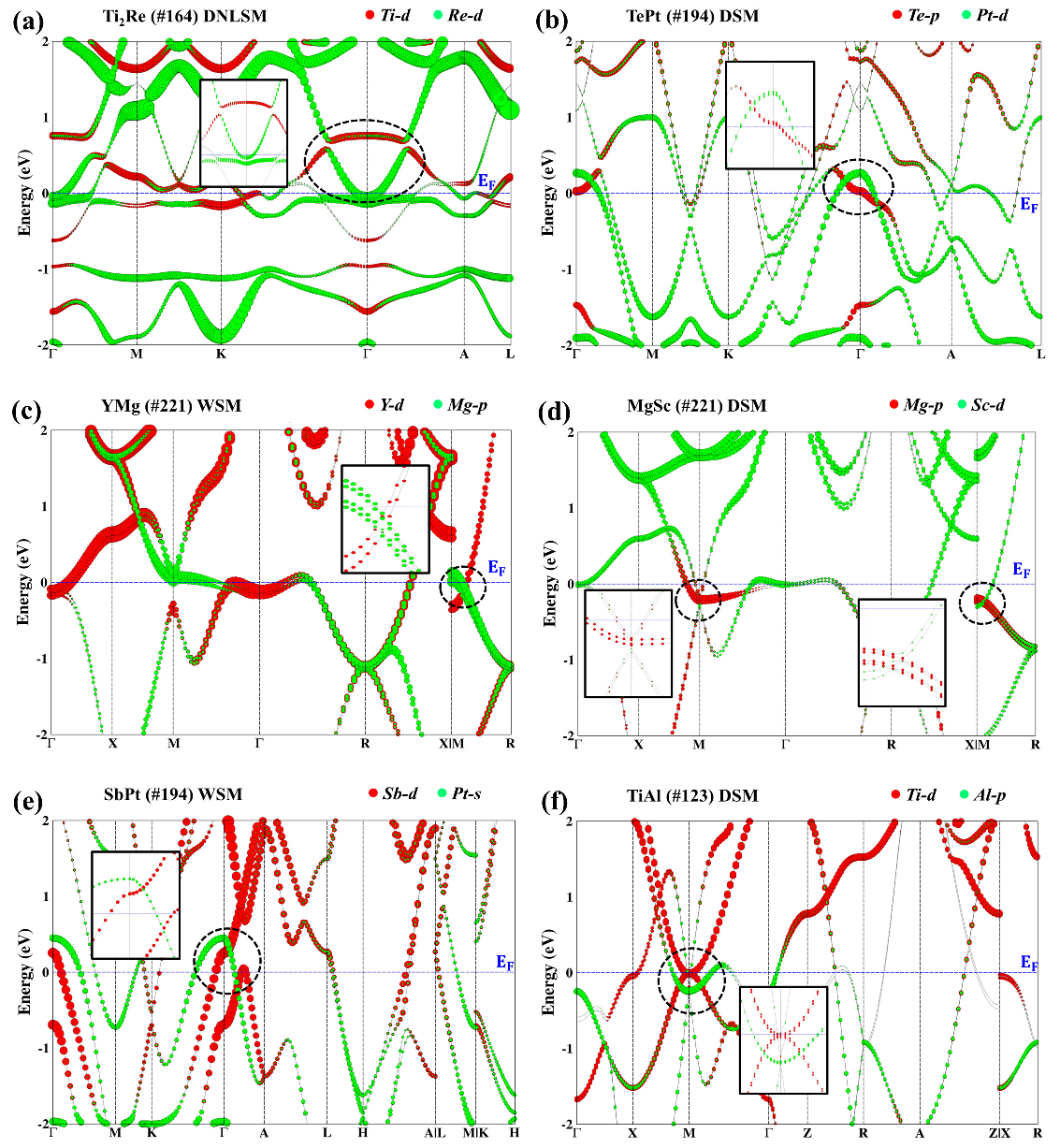}
	\caption{The projected band structures of some selected materials , $i.e.$, (a) Ti$_2$Re, (b) TePt, (c) YMg, (d) MgSc, (e) SbPt, (f) TiAl, from the predicted RF datapoints.}
	\label{fig:enter-label}
\end{figure}		
     We have also compared the outcome datapoints of the SVM and RF models and there are 22 such materials that are common in the prediction spaces of both models (Table S6). We have analyzed 19 common materials (excluding the radioactive materials) using $first-principles$ calculations and their topological nature is listed in Table VI. Hence, these materials have the highest probability of a topologically non-trivial phase. The crystal structures and the $first-principles$ calculations of some selected materials from table 6 are shown in Fig. 8 and 9, whereas their k path for band structure calculations are included in Fig. S5. Fig. S8 in the supplementary information can be refereed for the analysis of the rest of the commonly predicted systems\cite{20}. 
		
		\begin{table}[htbp]
			\centering
			\caption{The identified topological phases of the commonly predicted materials with SVM and RF models using $first-principles$ calculations.}
			\begin{tabular}{|c|c|c|c|c|c|}
				\hline
				\textbf{Material} & \textbf{SG No.} & \textbf{Phase} & \textbf{Material} & \textbf{SG No.} & \textbf{Phase} \\
				\hline
				SnPd$_3$     & 221 & DSM   & TePd        & 194 & WSM \\
				SbPt$_7$     & 225 & DSM   & TiZn$_2$    & 194 & WSM \\
				GaPd$_2$     & 62  & DNLSM & GaPd        & 198 & WSM \\
				BPd$_6$      & 70  & WSM   & Te$_3$Pd$_{10}$ & 216 & WSM \\
				SiPd$_2$     & 189 & WSM   & Pd$_3$Pb    & 221 & WSM \\
				ScZn$_{12}$  & 139 & WSM   & AuPd$_3$    & 221 & WSM \\
				P$_3$Pd$_7$  & 146 & WSM   & BeZn$_{13}$ & 226 & WSM \\
				GePd$_2$     & 189 & WSM   & Ge$_8$Pd$_{21}$ & 88  & WSM \\
				GePt$_2$     & 189 & WSM   & AsPd$_2$    & 189 & WSM \\
				\hline
			\end{tabular}
			\label{tab:svm_rf_common_materials}
		\end{table}
		\begin{figure}[h]
			\centering
			\includegraphics[width=0.9\linewidth]{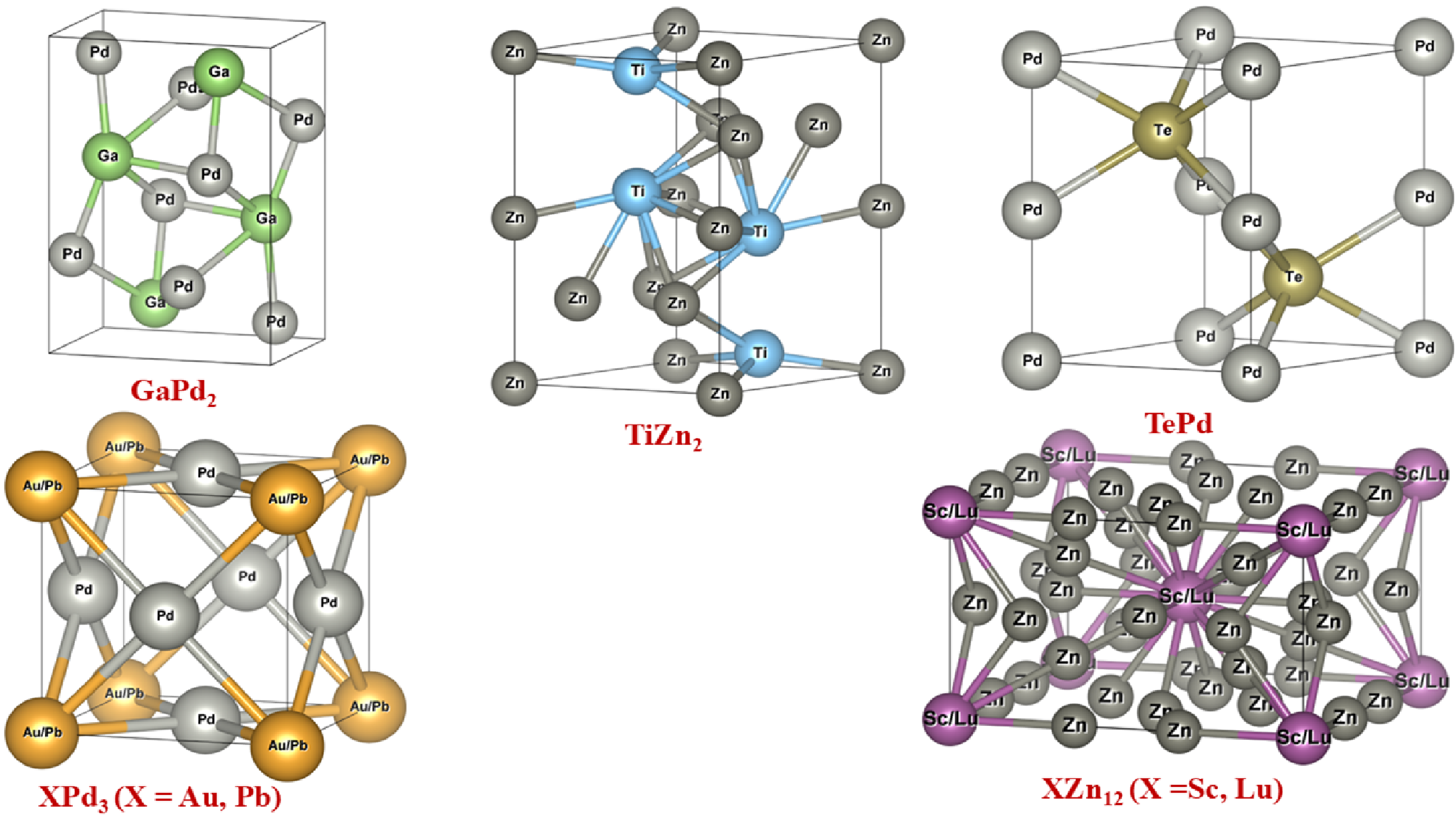}\\[1ex]
			\includegraphics[width=0.9\linewidth]{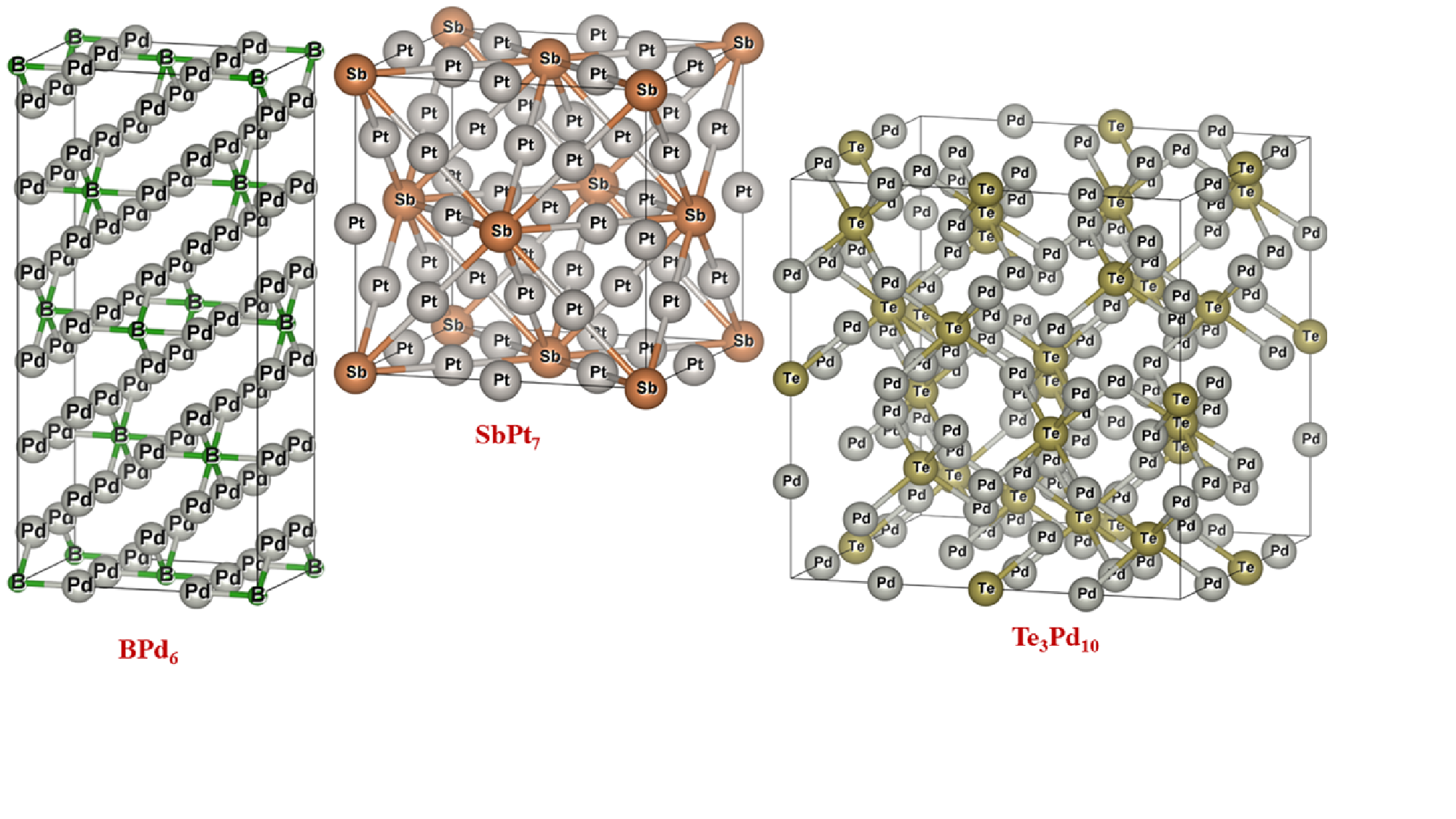}
			\caption{The crystal structures of some selected materials from the commonly predicted SVM and RF datapoints.}
			\label{fig:common_datapoints}
		\end{figure}
		
			\begin{figure}
			\centering
			\includegraphics[width=1\linewidth]{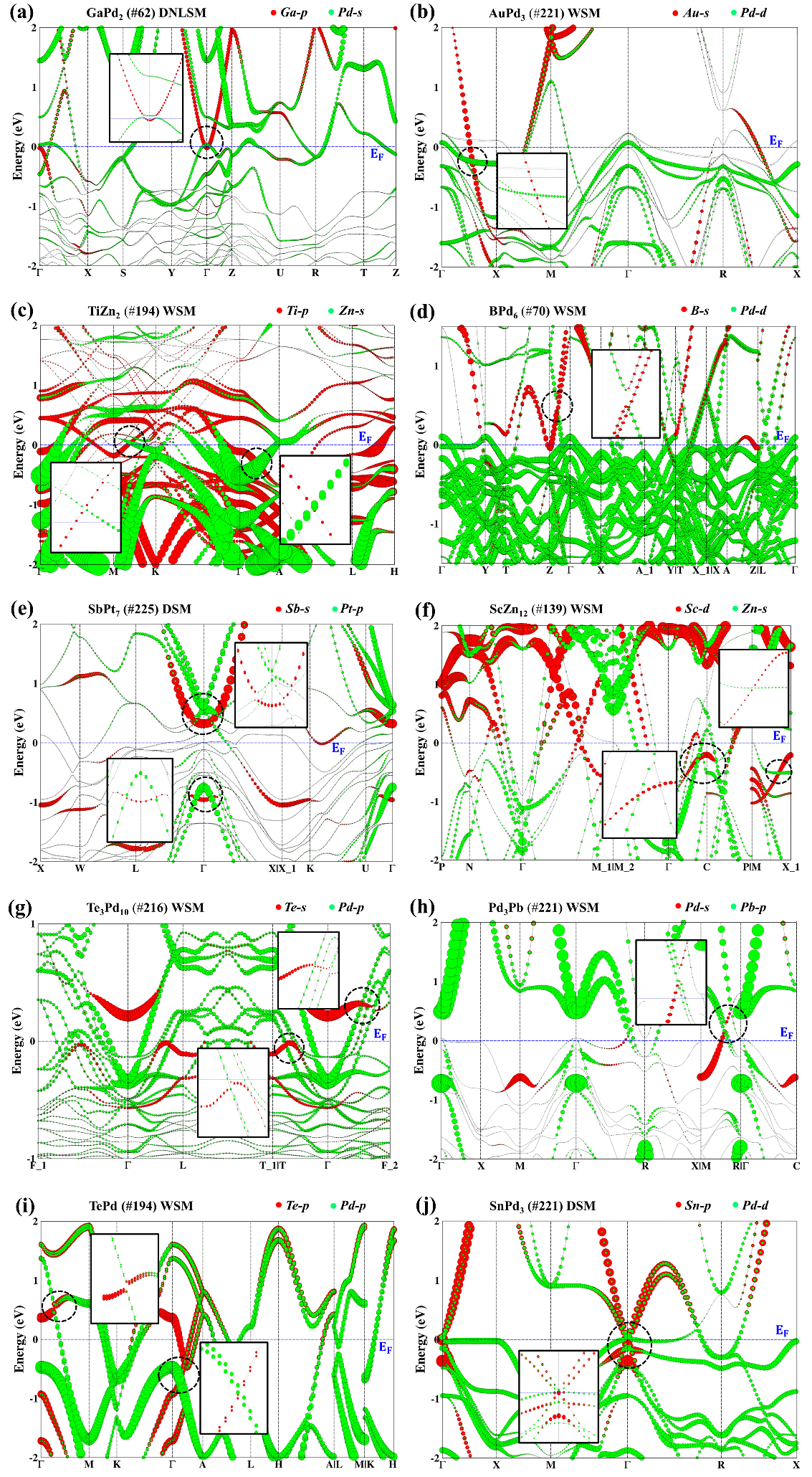}
			\caption{The projected band structures of some selected materials , $i.e.$, (a) GaPd$_2$, (b) AuPd$_2$, (c) TlZn$_2$, (d) BPd$_6$, (e) SbPt$_7$, (f) ScZn$_{12}$, (g) Te$_3$Pd$_{10}$, (h) LuZn$_{12}$, (i) Pd$_3$Pb, (j) TePd, from the commonly predicted SVM and RF datapoints.}
			\label{fig:enter-label}
		\end{figure}		
Within the given limitations in the accuracy of our models, we could enhance the number of predicted topological materials by selecting a less conservative baseline for predictions, and this approach would allow us to increase the count of materials identified as TMs with an increased value of recall, but at the cost of a decrease in the value of precision. This method of feature engineering is reliable for the identification and discovery of novel TMs with efficient accuracy. However, there may be noise in the labelling of our training data, and these noises may cause differences in our results, and if they are refined, the result of classification may differ. Moreover, the ground state predictions of the materials highly depend on the choice of exchange and correlation functionals in the first principles calculations. Our results in Tables IV, V, VI, and Figs. 5, 7, 9 correspond to the GGA-PBE functional with inclusion of SOC; hence, these results may have changes with the applicability of other functional approaches defined in the DFT. 
\section{\label{sec:level3}LIMITATIONS\protect}
Although both ML models have shown promising results for classifications of TMs, there are some limitations to the current method that need to be discussed. The main problem comes from the available dataset and the way the dataset is set up. The limited availability of the dataset is a significant constraint for the study. The original data had a very uneven class distribution, with 6098 trivial materials ($\mathrm{G_1}$) greatly outnumbering 1376 non-trivial topological materials ($\mathrm{G_2}$). In addition, the dataset contained many space groups as well as diverse chemical element combinations, which makes it harder to infer the pattern in the data. We used random undersampling to fix this imbalance and lessen its effect on model training. This change was necessary to keep bias out of the classification boundaries, but it also made the trivial class in the training set less diverse. This reduction creates a limitation in representation because the model is trained on a smaller view of the materials from a trivial space. So, even though the overall accuracy is high, the model might not perform well when it is implemented on different types of imbalanced real-world datasets. This could make the number of false positives go up. This trade-off between balance and diversity is a serious problem in ML for materials science. 
\section{\label{sec:level3}CONCLUSION\protect}
In this study, we have used a frequency-based statistical approach for the creation of features to effectively implement machine learning techniques in the classification and discovery of non-trivial topological materials. The effectiveness of the suggested approach has been demonstrated by the classification accuracies of 82\% and 83\% for Support Vector Machine (SVM) and Random Forest (RF) models, respectively. We identified 22 common materials predicted by both models, which were further confirmed by density functional theory calculations. As an interpretable and computationally cheap substitute for traditional symmetry-based or structural-property-dependent descriptors, the frequency-based descriptors are derived solely from elemental occurrences. The models demonstrated dependable generalization, as confirmed by cross-validation and $first-principles$ computations, despite the small dataset. However, prediction accuracy is limited when material-specific physical or electronic properties are not included in the feature space. Model performance and generalizability could be significantly increased by expanding the feature space with more material descriptors, such as orbital characteristics, formation energy, or quantities derived from band structures. These findings confirmed that a statistical frequency-based method offers a promising avenue for speeding up the identification of candidate topological materials when combined with machine learning and filtered using domain-specific criteria. To further increase the predictive ability of machine learning models for the discovery of quantum materials, future research may investigate hybrid feature spaces that combine statistical, structural, and electronic data.				
		\begin{acknowledgments}
		All the authors acknowledge National Supercomputing Mission (NSM) for providing computing resources of ‘PARAM Siddhi-AI’, under National PARAM Supercomputing Facility (NPSF), C-DAC, Pune and supported by the Ministry of Electronics and Information Technology (MeitY) and Department of Science and Technology (DST), Government of India. One of the authors (R. K.) would like to thank the Council of Scientific and Industrial Research (CSIR), Delhi, for financial support. The corresponding author (M.S.) wants to acknowledge the financial support by “Anusandhan National Research Foundation (ANRF)” [formerly known as Science and Engineering Research Board (SERB)], Department of Science and Technology (DST), Government of India, under Grant No. EEQ/2023/000913.
		\begin{description}
			\item[Author contributions]
			These authors contributed equally to this work.
			\item[Conflict of interest statements]
			The authors declare no competing financial interest.
		\end{description}
			\end{acknowledgments}
\section*{Data availability}
The codes and data supporting this work are available at:\\
\url{https://github.com/Zodinpuia11-sys/Frequency_based_TM_classification}

	\onecolumngrid	
\bigskip
	\centering
\noindent\textbf{References}
	\twocolumngrid

	%
	
\end{document}